\documentclass[aps,pra,preprint,superscriptaddress]{revtex4-2}   	

\usepackage{graphicx}
\usepackage{amssymb}
\usepackage{amsmath}
\usepackage{multirow}
\usepackage{braket}

\usepackage{array}
\newcolumntype{P}[1]{>{\centering\arraybackslash}p{#1}}

\begin{document}

\title{Anomalies in the rotational spectra of $^{86}$Sr ULRRM dimers}

\author{C. Wang}
\author{Y. Lu}
\author{A. Li}
\author{S. K. Kanungo}
\author{T. C. Killian}
\author{F. B. Dunning}
\affiliation{Department of Physics and Astronomy, Rice University, Houston, TX  77005-1892, USA}
\author{S. Yoshida}
\affiliation{Institute for Theoretical Physics, Vienna University of Technology, A-1040 Vienna, Austria, EU}

\begin{abstract}
Anomalies in the rotational structure of $^{86}$Sr $^3S_1$ dimer ultralong-range Rydberg molecules (ULRRMs) created in a cold strontium gas by two-photon excitation via the intermediate $5s5p~^3P_1$ state are reported.  Measurements reveal that the distribution of product rotational states is sensitive to intermediate state detuning.  Comparative studies using $^{84}$Sr $^1S_0$ and $^3S_1$, and $^{86}$Sr $^1S_0$ dimers display no similar behavior, indicating that the observed behavior is peculiar to $^{86}$Sr triplet dimers.  While we have no definitive hypothesis as to the physical mechanism responsible for this behavior, possible explanations might involve the very different scattering lengths for $^{84}$Sr and $^{86}$Sr, or the interchange of spin and rotational angular momentum.
\end{abstract}

\maketitle

\section{Introduction}
Ultralong-range Rydberg molecules (ULRRMs) comprise a Rydberg atom in whose electron cloud is embedded one (or more) ground state atoms that are weakly bound through their scattering of the Rydberg electron.  Such scattering is frequently described using a Fermi pseudo-potential resulting in a molecular Born-Oppenheimer potential that can support multiple vibrational levels
~\cite{dky24,srs18,fhs20}.  
In the case of strontium $ns$ Rydberg states, the ground $v=0$ vibrational state is of particular interest because it is strongly localized near the outer classical turning point at an internuclear separation, $R_n$, of $\sim 1.8(n-\delta)^2 a_0$ where $n$ is the principal quantum number, $\delta$ the quantum defect, and $a_0$ the Bohr radius.  This has, for example, enabled ULRRMs to be used as a `ruler' to explore, for example, correlation functions
~\cite{wkd19,kld23}.

The vibrational structure of ULRRMs has been the subject of several studies using cold gases of both alkali and alkaline earth atoms
~\cite{bbn09,bbn10,csw18}.    
More recently, however, spectroscopic measurements have been extended using strontium to investigate their rotational structure
~\cite{lwk22,wlk24}.
These studies focused on the production of both $^{84}$Sr and $^{86}$Sr dimers.  $^{86}$Sr is of particular interest because the $^{86}$Sr-$^{86}$Sr atom-atom $s$-wave scattering length, $a_s$, is unusually large, $a_s=811a_0$, and is comparable in size to the internuclear separation for $v=0$ ULRRM dimers with values of $n\sim25$.  In contrast, the $s$-wave scattering length for $^{84}$Sr-$^{84}$Sr collisions is only $a_s\sim123 a_0$.  Thus comparisons between measurements with $^{84}$Sr and $^{86}$Sr can provide insights into the role played by the initial atom-atom interaction in dimer formation.

\begin{figure}
    \centering           \includegraphics[width=0.6\linewidth]{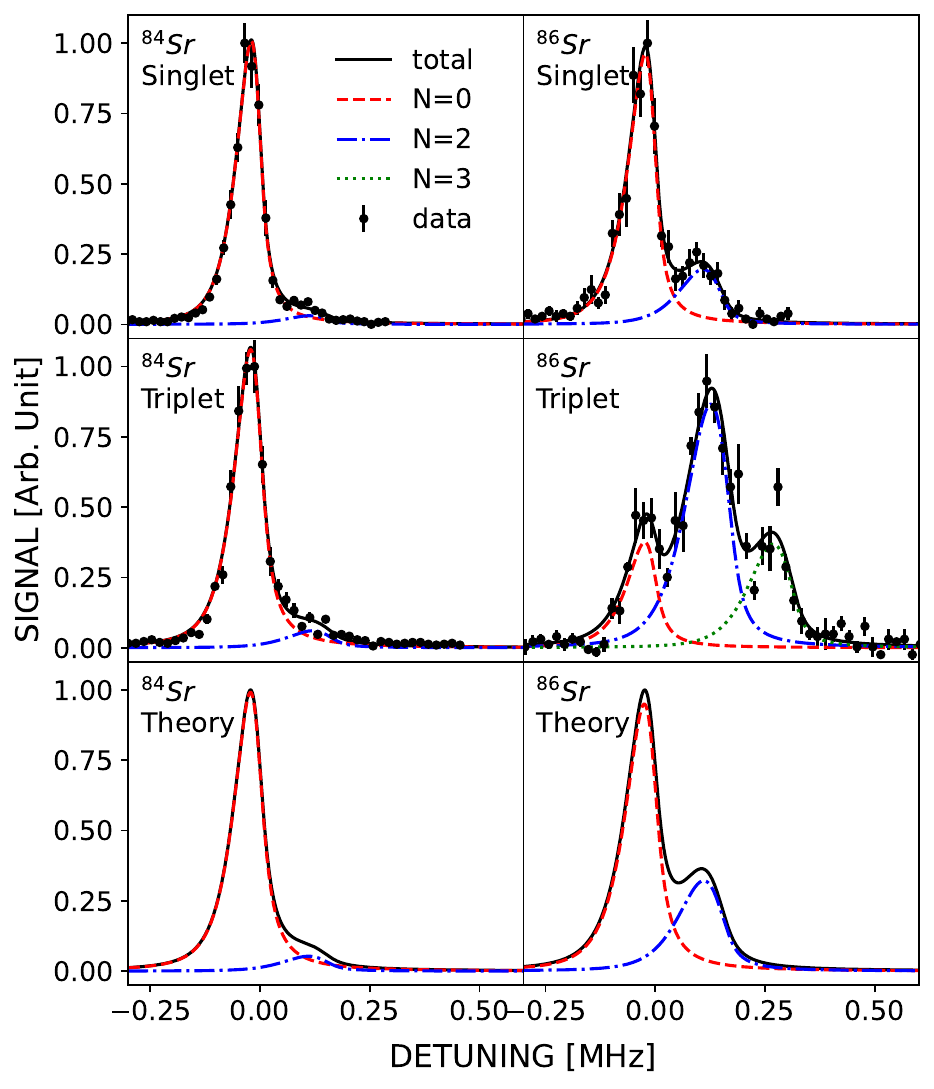}
    \caption{\label{fig:bothisotopes}
    Comparison of photoassociation spectra measured for $^{84}$Sr and $^{86}$Sr singlet and triplet dimers.  The data were recorded using similar trap operating conditions and densities. The results are fit by contributions from the rotational, i.e., $N$, states indicated and using the theoretically-predicted lineshapes obtained with the model discussed in Section ~\ref{sec:theory}.  Also included for comparison are theoretical predictions using the same model.  
    }
\end{figure}

Initial studies of the formation of rotationally-excited dimers (in the ground $v=0$ rotational state) through photoassociation focused on $^{86}$Sr 5sns $^3S_1$ Rydberg states, created by two-photon excitation via the intermediate $5s5p~^3P_1$ state
~\cite{lwk22}.
Measurements were subsequently extended to $^{84}$Sr~5sns~$^1S_0$ dimers created by two-photon excitation via the intermediate 5s5p~$^1P_1$ state
~\cite{wlk24}.  
These studies highlighted the roles played by the scattering length, photon momentum transfer, and sample temperature in controlling dimer formation.  Representative measured photoassociation spectra obtained using both isotopes are presented in Fig.~\ref{fig:bothisotopes}. 
   As is evident from Fig.~\ref{fig:bothisotopes}, 
 in the case of $^{84}$Sr singlet and triplet dimers, the measured profiles can be well fit by contributions from just two rotational states, as predicted by theory.  While this is also true of $^{86}$Sr singlet dimers, the $^{86}$Sr triplet spectrum differs markedly from the others, displaying three, not two, well-defined peaks.  Here we describe new measurements designed to probe the possible origins of this very different behavior, which is peculiar to $^{86}$Sr triplet dimers.

\section{Experimental Method}
\begin{figure}
        \centering
        \includegraphics[width=0.5\linewidth]{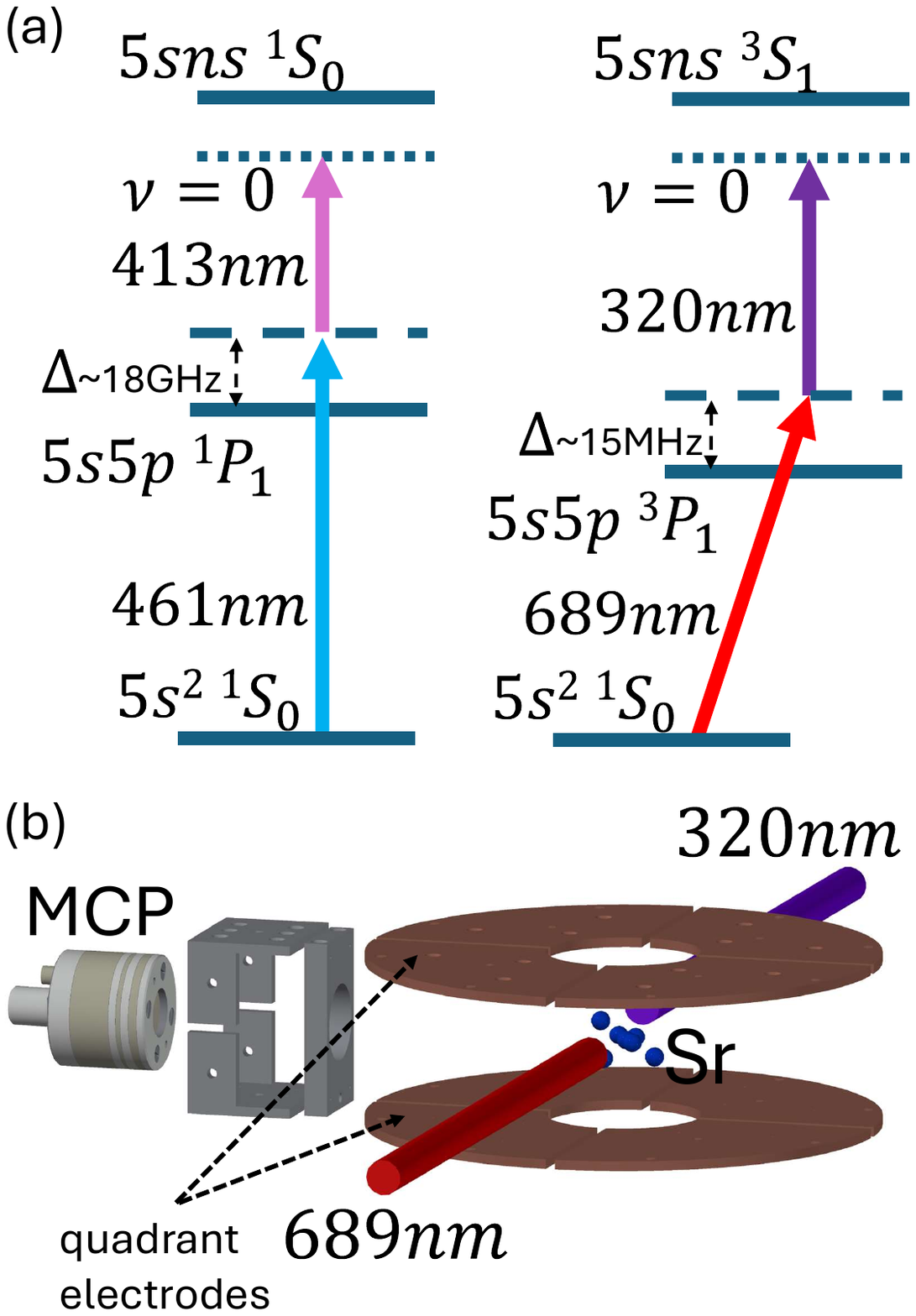}
        \caption{\label{fig:schematic}
     (a) Photoexcitation schemes used to create the $^{84}$Sr and $^{86}$Sr singlet and triplet dimers. (b) Schematic diagram of the apparatus.
     }
 \end{figure}

The present experimental apparatus is shown in Fig.~\ref{fig:schematic} and has been described earlier\cite{ssk14,dmy09}.  
Briefly, strontium atoms are laser-cooled to a few microkelvin and loaded into a `pancake'-shaped optical dipole trap (ODT) formed from two crossed elliptical 1064~nm laser beams.  Evaporative cooling is used to further lower and control the atom temperature.  Triplet Rydberg dimers are created by two-photon excitation via the intermediate 5s5p $^3P_1$ state.  The required radiation at 689 and 320~nm is provided by diode laser systems that are stabilized to high-finesse ultralow-expansion (ULE) cavities.  Singlet dimers are produced by two-photon excitation via the $^1P_1$ intermediate state, the frequencies of the required 461 and 413~nm lasers again being stabilized to a ULE cavity.  To limit trap loss and heating, the 689~nm (461~nm) lasers are blue detuned from the intermediate $^3P_1$ ($^1P_1$) levels.  Photoassociation spectra are recorded by tuning the 320~nm (413~nm) laser while maintaining the intermediate state detuning constant.  To minimize the net photon momentum transfer (and Doppler broadening) the lasers counter-propagate.  Laser pulse durations of $\sim20~\mu$s are employed, which, when taking into account transform broadening and laser linewidths, amounts to an overall effective linewidth of $\sim 70$~kHz.  This linewidth is comparable to that used in earlier $^{84}$Sr $n^1S_0$ measurements but is significantly less than that employed in earlier $^{86}$Sr $n^3S_1$ measurements, $\sim 120-140$~kHz, which enabled the present, more detailed, spectroscopic studies.  The ODT is loaded with $\sim 8-15\times 10^5$ atoms, resulting in trap densities of up to $\sim1\times 10^{13}$~cm$^{-3}$.  The final atom number and temperature are determined through absorption imaging on the 5s$^2~^1S_0 \rightarrow$ 5s5p~$^1P_1$ transition following release of the atoms from the ODT and a fall time of $\sim25$~ms.

The number of molecules formed is measured through ionization in a pulsed electric field
~\cite{sdu83,gal94}.  
The resulting electrons are directed to a dual-microchannel-plate detector for recording.  Experimental limitations on the size of the field that can be applied in the experimental volume restricted measurements to states with $n\geq29$.  Excitation rates were kept low, typically $\leq 0.3$ per experimental cycle, to minimize possible effects due to Rydberg-Rydberg interactions.  The need to maintain a constant sample temperature and density during a set of measurements limited the number of experimental cycles that could be undertaken using a single cold-atom sample to $\sim200$.

\section{Theoretical Analysis}
\label{sec:theory}
The present results are analyzed using a theoretical model that has been described in detail previously
~\cite{lwk22} 
and that takes into account the initial ground-state atom-atom interaction, photon momentum transfer, and sample temperature.

Briefly, a partial-wave representation $\vert k,N^\prime, M^\prime_N\rangle$ is adopted for the relative motion of the colliding atom pair where $\hbar k$ is the relative momentum of the colliding pair, $N^\prime$ is the rotational quantum number, and $M^\prime_N$ its projection on the quantization axis.
For an initial scattering state in which both (bosonic) atoms are in the same, $5s^2$, ground state, symmetry allows only even waves for the initial scattering state when, if the effects of photon recoil momentum are small, only even $N=0, 2, 4\cdots$ rotational states in the Rydberg dimer should be created.  However, in the formation of an ULRRM dimer one atom in the initial ground-state atom pair acquires photon momentum during its excitation which, given the large internuclear spacing in the dimer, can also lead to significant rotational angular momentum transfer (included in the model) to the atom pair, i.e., the dimer, allowing formation of odd-$N$ rotational states.

In addition, since the total orbital angular momentum of the valence electrons in strontium vanishes in the Rydberg states used here then, neglecting spin, the total mechanical angular momentum of the product dimer is characterized by $N$ and $M_N$.  The vibrational and rotational part of the transition amplitude is given by an inelastic form factor that can be written
\begin{eqnarray}
\label{eq:FranckCondon}
    &&F_{v,N,M_{N}}(k,N^{\prime},M_N^{\prime})
    \nonumber\\
    &&=\frac{1}{\sqrt{2}}\langle v,N,M_N\vert[e^{-i{\vec{\kappa}/2{\cdot\vec{R}}}}+(-1)^N e^{i(\vec{\kappa}/2)\cdot\vec{R}}]\vert k,N^{\prime},M^{\prime}_N\rangle
    \nonumber\\
\end{eqnarray}
where $v$ denotes the final vibrational state and $\vec\kappa$ is the sum of the wave vectors associated with the laser fields.  (Since the separation between the $v=0$ and $v=1$ vibrational levels is large, it is assumed that there is no mixing between vibrational and rotational motions.)  This form factor resembles the Franck-Condon (FC) overlap important in molecular spectroscopy but now including the effects of photon recoil momentum.  This form factor is evaluated using spherical representations of both the initial scattering state of the atomic pair, and the final Rydberg dimer state.

The overall final excitation strength, $f(\omega)$, is obtained by averaging over the Boltzmann distribution of the relative momenta of atoms in the trap incorporating the effective laser linewidths, and enforcing energy conservation.
\begin{eqnarray}
    f(\omega)
    &\propto&\sqrt{\frac{2}{\pi}}\left(\frac{\hbar^2}{\mu k_BT}\right)^{3/2}
    \sum_{v,N,M_N}\int dk k^2 e^{-\hbar^2k^2/(2\mu k_BT)}
    \nonumber\\
    &&
    \times\sum_{N^\prime,M^\prime_N}A_N\vert F_{v,N,M_N}{(k,N^{\prime},
    M_N^{\prime}})\vert^2 L_{v,N}(k,\omega),
    \label{eq:final strength}
\end{eqnarray}
where $\mu=m/2$ is the reduced mass and we have introduced $A_N$ as a phenomenological fitting parameter modeling the contribution to  the spectrum associated with each value of $N$.  $T$ is the sample temperature, and $L_{v,N}(k,\omega)$ is the 
normalized line-shape function. $L_{\nu,N}$ can be well represented by a distribution that includes the Doppler broadening induced by the center-of-mass (CM) momentum $\hbar\vec{k}_{CM}$ of the dimer.
\begin{eqnarray}
&&L_{v,N}(k,\omega)
\nonumber\\
&&=\frac{1}{\pi}\left(\frac{\hbar^2}{2\pi M k_BT}\right)^{3/2}
\int d\vec{k}_{\rm CM} e^{-\hbar^2k_{\rm CM}^2/(2 M k_BT)}
\nonumber\\
&&\times\frac{2\Gamma}
{4(\omega - \hbar\vec{\kappa}\cdot\vec{k}_{\rm CM}
/M + \hbar k^2/(2\mu) - E_{v,N}/\hbar)^2 + \Gamma^2}
\label{eq:Lorentzian}
\end{eqnarray}
where $M$ ($=2m$)is the total mass  of the dimer, $\hbar^2k^2/2\mu$ is the initial relative kinetic energy of the collision pair, $\omega$ the two photon detuning from atomic resonance
for an atom at rest, and $\Gamma$ the full-width-at-half-maximum (FWHM) linewidth.  The peak positions in Eq.~\ref{eq:Lorentzian} are determined approximately by the initial relative kinetic energy of the collision pair, i.e., $E_{v,N} =\hbar\omega + \hbar^2k^2/2\mu$ on resonance, and energy conservation
during excitation.  The thermal averaging inherent in Eqs.~\ref{eq:final strength} and ~\ref{eq:Lorentzian} leads to an increased linewidth and asymmetric non-Lorentzian line profiles.  The photoassociation profile can be calculated using the eigenenegies $E_{\nu,N}$ obtained by numerically diagonalizing the molecular potential.

The Franck-Condon factor (Eq.~\ref{eq:FranckCondon}) is also evaluated including the effects of $s$-wave scattering lengths
\cite{lwk22}.  In the calculations the phenomenological parameter is always set to $A_N=1$.  On the other hand, in order to extract rotational energies $E_{v=0,N}$ and the contribution from each rotational level to the measured spectra, each is fit using Eqs.~\ref{eq:final strength} and \ref{eq:Lorentzian}, with $A_N$, $\Gamma$, and $E_{v,N}$ treated as adjustable parameters.

\section{Results and Discussion}
\begin{figure}
        \centering
        \includegraphics[width=0.6\linewidth]{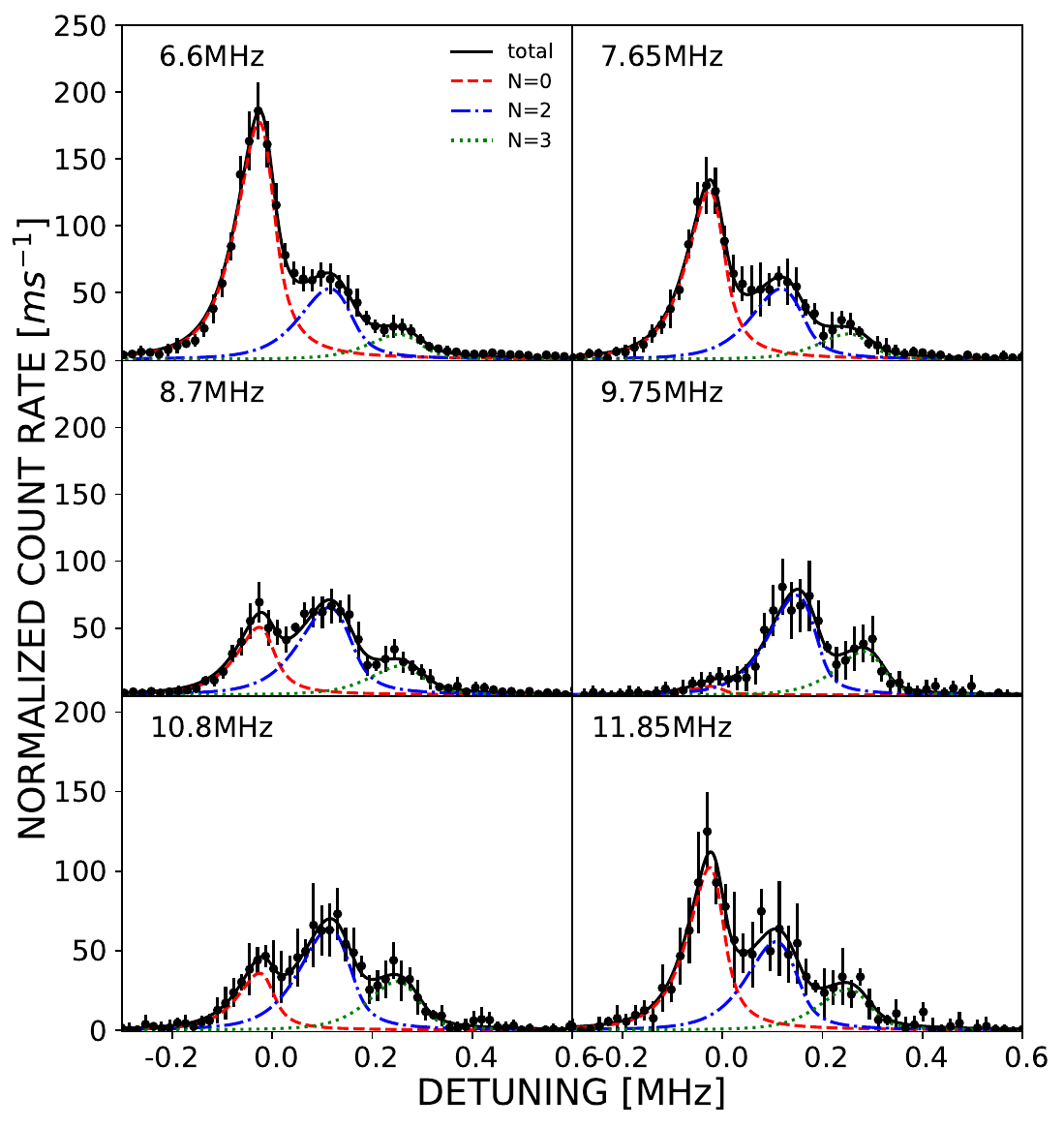}
        \caption{
    \label{fig:spectra}
    Photoassociation spectra recorded when creating $^{86}$Sr~29$^3S_1$ dimers.  The different detunings of the 689~nm laser from the  $^3P_1$ intermediate state are indicated in each panel.  The trap operating conditions are maintained constant and the spectra normalized for any small changes in laser powers or atom density as described in the text.  The sample temperature is 1.2~$\mu$K.  The solid lines show fits to the data assuming contributions from $N=0,~2$, and 3 rotational states and the theoretically-predicted lineshapes (see text).  (The $N=3$ profile was taken from calculations for higher sample temperatures.) The zero of the frequency axis in each figure is set to the position of the contribution from atoms with no relative or center-of-mass momentum.
    }
\end{figure}

\begin{figure}
        \centering
        \includegraphics[width=0.6\linewidth]{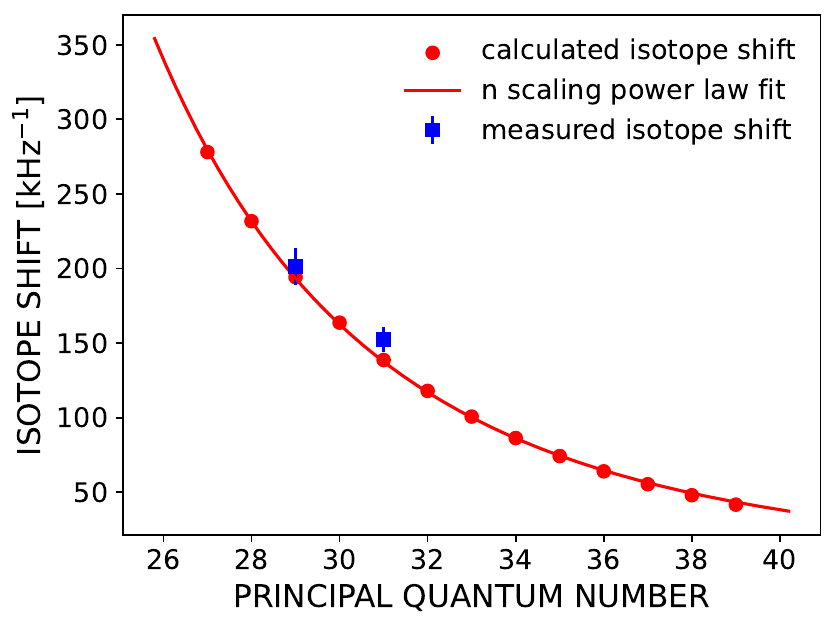}
        \caption{\label{fig:isotope shifts}
    Measured and calculated isotope shifts for ($v=0$) $^{84}$Sr and $^{86}$Sr dimers referenced to that for $^{84}$Sr dimers. 
    }
\end{figure}

To gain new insights into the origins of the anomalous behavior observed with $29^3S_1$ dimers, a series of spectra for excitation of $^{86}$Sr ULRRM ground-vibrational-state dimers were recorded as a function of intermediate state detuning while maintaining the trap operating conditions and laser powers near constant, and the results normalized for any small changes in laser powers or trap density.  The total signal counts recorded at a given two-photon detuning are proportional to the total laser exposure time ($T_{expo}$), the product of the number of atoms, $N$, and their peak density, $n_0$, the product of the excitation laser intensities, $I_{beam1}$ and $I_{beam2}$, and the inverse square of the intermediate state detuning, $\Delta$,
\begin{equation}
    S\propto T_{expo}Nn_0\frac{I_{beam1}I_{beam2}}{4\Delta^2}.
\end{equation}
In the plots the data are normalized by these factors and we report the rate ($S/T_{expo}$) that would scale to a sample of $N=10^6$ atoms, a peak density $n_0 = 10^{12}$~cm$^{-3}$, laser intensities of 1~mW/cm$^2$, and intermediate state detuning of $\Delta/2\pi = 1$~MHz.  The signal rate is also proportional to the detector efficiency which we assume is constant and is not included in the normalization.
The resulting spectra are shown in Fig.~\ref{fig:spectra},
each of which contains three features which, based on earlier observations, we assign to the creation of $N=0,~2,$ and 3 rotational states.  This assignment was reinforced by making measurements of the isotope shifts in the binding energy of the lowest energy features in the $^{84}$Sr and $^{86}$Sr dimer spectra and comparing these to theoretical predictions.  These shifts were determined by first tuning the 320~nm laser to the atomic line and then determining the detuning required to excite the lowest-energy feature in each dimer spectrum from the fit values $E_{v,N}$.  The difference in the measured detunings for $^{84}$Sr and $^{86}$Sr was then compared to theorectical predictions.  As shown in Fig.~\ref{fig:isotope shifts}, the measured differences agree well with theoretical predictions and
indicate that the feature labeled $N=0$ in the $^{86}$Sr dimer spectrum does indeed result from formation of $N=0$ dimers.  Further support for this assignment is provided by measurements of the peak separations.  The average separation between the  $N=0$ and 2 peaks in the triplet spectra determined from the fit values $E_{v,N}$ is $\sim186$~kHz, which is consistent with the separations between the $N=0$ and 2 states measured in earlier studies
~\cite{wlk24}.  
In addition, the separation between the $N=0$ and $N=3$ features, $\sim 324$~kHz, is also consistent with this attribution.

\begin{figure}
        \centering
        \includegraphics[width=0.6\linewidth]{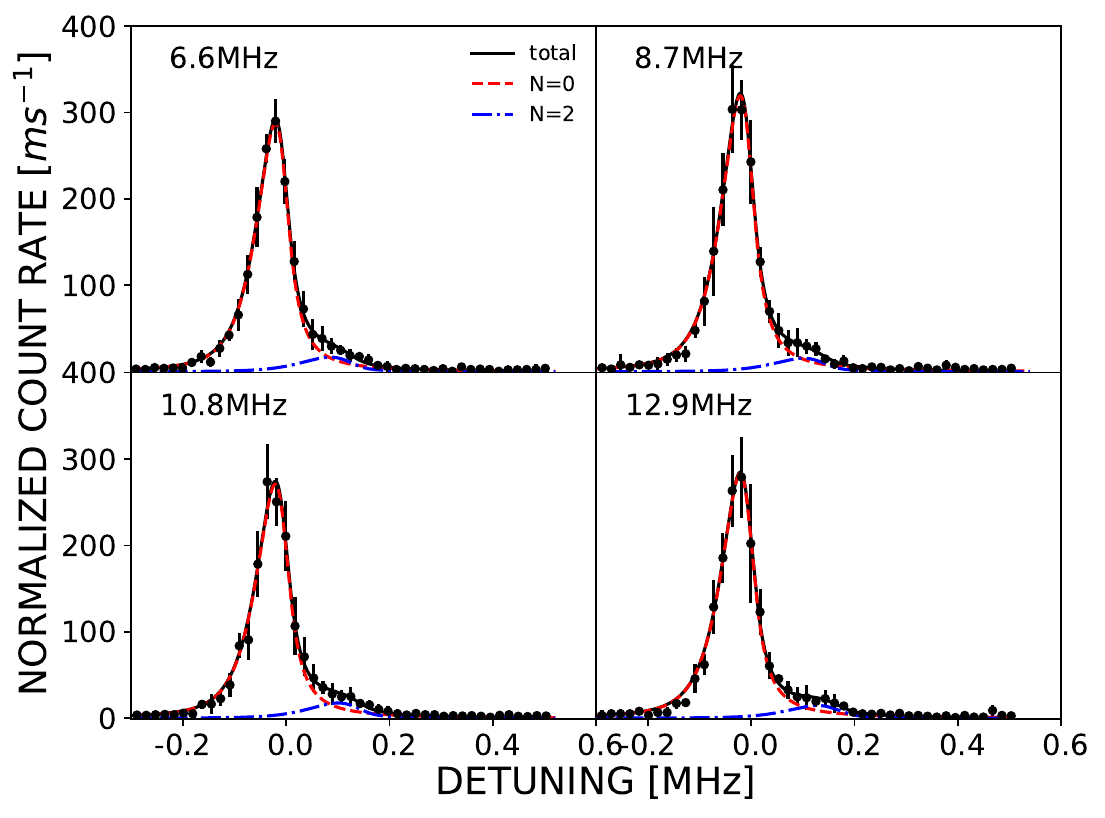}
        \caption{
    \label{fig:detunings}
    Photoattachment spectra measured when creating $^{84}$Sr $29^3S_1$ dimers with the different intermediate state detunings indicated.  The spectra were recorded in zero magnetic field using identical trap operating conditions and are normalized for small changes in laser intensities and atom densities.  Each profile can be well fit by assuming contributions from just $N=0$ and 2 rotational states.
    }
\end{figure}

Interestingly, as seen in Fig.~\ref{fig:spectra}, the size of the triplet $^{86}$Sr~$N=0$ feature varies dramatically with intermediate detuning, whereas the $N=2$ and $N=3$ features show relatively small variations.  For comparison, a series of measurements of $^{84}$Sr~29$^3S_1$ triplet dimer formation were undertaken for a number of different intermediate detunings.  The results are presented in Fig.~\ref{fig:detunings}.   
In contrast to the $^{86}$Sr~$^3S_1$ observations, the spectra contain a single dominant $N=0$ feature that does not display any strong oscillatory dependence on intermediate state detuning.  Indeed, the spectra are very similar to those seen when creating $^{84}$Sr $5s29s~^1S_0$ dimers
~\cite{wlk24}. 

\begin{figure}
            \centering
            \includegraphics[width=0.6\linewidth]{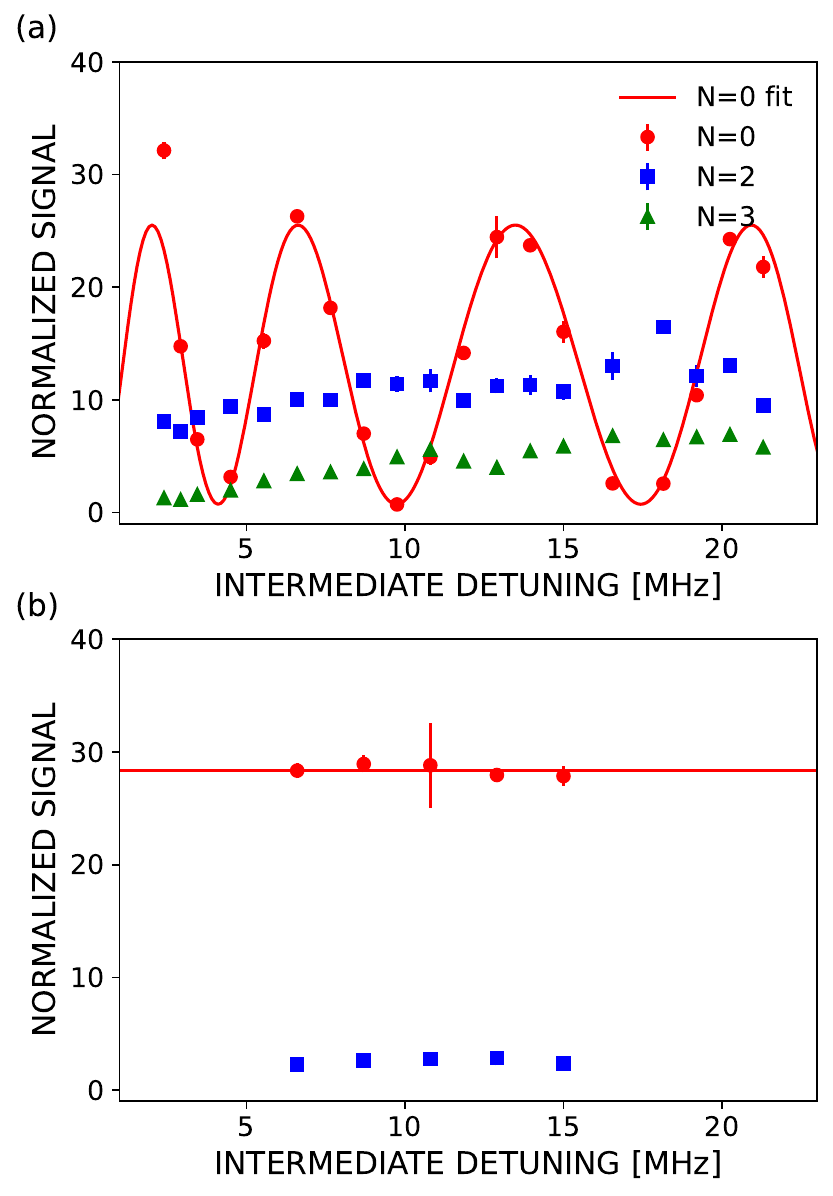}
                \caption{\label{fig:integrated}
        (a) Integrated signals associated with the $N=0, 2,$~and 3 features as a function of intermediate state detuning for $^{86}$Sr 29$^3S_1$.  (b) Measurements of the total $N=0$ and $N=2$ dimer production when creating $^{84}$Sr~29$^3S_1$ dimers.  The two data sets are each normalized for differences in trap densities and laser intensities.
    }
\end{figure}

\begin{figure}
         \centering
        \includegraphics[width=0.5\linewidth]{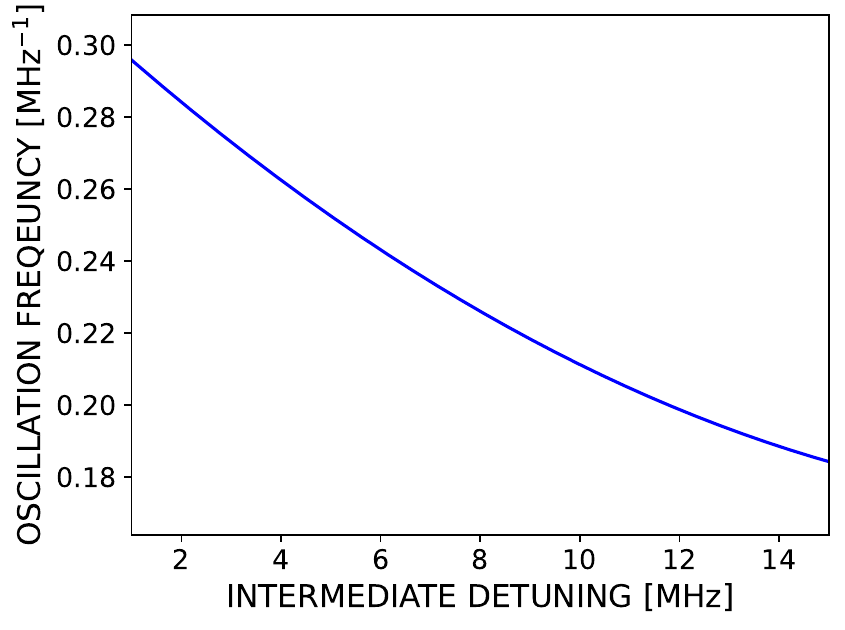}
          \caption{\label{fig:values}
    Values of the oscillation "frequency" $f$ measured as a function of intermediate state detuning (see text) for $^{86}$Sr 29$^3S_1$ dimers.
    }
\end{figure}

The differences between the $^{84}$Sr and $^{86}$Sr spectra are highlighted in Fig.~\ref{fig:integrated} 
which shows the integrated areas of the $N=0$, 2, and 3 features in the $^{86}$Sr spectra together with the total $^{84}$Sr dimer signals as a function of 689~nm laser detuning (with the signals normalized as described earlier).  The $^{86}$Sr $N=0$ signal displays a large-amplitude oscillation that can be well-fit by a sinusoidal function of the form
\begin{equation}
S=A\sin2\pi[\Delta(f_0 + k_1\Delta +k_2\Delta^2) + \phi]
\end{equation}
where the amplitude $A$ and phase constant $\phi$ are constants, $\Delta$ is the intermediate state detuning, and $f_0$, $k_1$ and $k_2$ are constants derived from the fit.  Values of the effective "oscillation" frequency $f=(f_0 + k_1\Delta + k_2\Delta^2)$ for $n=29$ are shown in Fig.~\ref{fig:values} 
and decrease steadily with increasing detuning.  A small oscillation is seen in the $N=2$ population that is antiphase to that for the $N=0$ population.  Data recorded using 31$^3S_1$ dimers displayed similar behavior to that seen with $29^3S_1$ dimers, but with smaller values of $f$. The observed oscillations cannot be attributed to Rabi oscillations
~\cite{ghk89}.  
The measured dimer production rates point to Rabi frequencies $\lesssim 2$~kHz that are consistent with transition rates given by Fermi's golden rule.  

We do not have a hypothesis for the physical mechanism giving rise to the very different observed behaviors seen in Fig.~\ref{fig:spectra} and~\ref{fig:detunings}, but it is likely that the different $^{84}$Sr and $^{86}$Sr scattering lengths, $811~a_0$ and $123~a_0$, respectively, play a role.  The scattering length for $^{86}$Sr is comparable in size to the internuclear separation for $v=0$ ULRRM dimers with values of $n\sim25$.  Because the value of $a_s$ defines the position of a node in the $s$-wave component of the atom-atom scattering wave function, the probability of finding an initial atom pair with separations close to this value is small.  Thus, for values of $n$ near 25 the Franck-Condon factor overlap between the N,N$^{\prime}$ component in the initial scattering state and final ($v=0$) molecular state is reduced. This limits dimer formation via the $s$-wave channel, which is otherwise the dominant scattering channel at low temperatures and which, in the absence of photon momentum transfer, can only populate the ground ($N=0$) rotational state.  This suppression, in turn, allows the contributions from higher-partial-waves ($d$-wave), which can directly populate rotationally-excited ($N=2$) states, to be observed.  (The $p$-wave component of the initial scattering state vanishes due to symmetry).  However, because of the smaller $^{84}$Sr-$^{84}$Sr atom-atom scattering length, dimer formation via $s$-wave scattering is no longer suppressed and the spectrum is dominated by the formation of $N=0$ states, although a small shoulder is evident that is associated with formation of $N=2$ states through $d$-wave scattering, see Fig.~\ref{fig:detunings}.  (Since the lasers counter-propagate, little creation of $N=1$ states through photon momentum transfer is expected.)
 
As seen in Fig.~\ref{fig:bothisotopes}, in the case of $^{84}$Sr singlet and triplet dimer formation, theory and experiment agree well
~\cite{wlk24}. 
While the same is true for $^{86}$Sr singlet dimers, this is not the case for triplet dimers where the model does not predict the presence of any feature that oscillates sinusoidally, leaving its origins in question.

To gain further insights into the origins of this behavior, photoassociation spectra were measured in the presence of a weak (a few Gauss) magnetic field applied to allow selective excitation of $29^3S_1$($m=1$ or -1) dimers.  Again, dramatic changes in the spectral profiles were seen as the 689~nm laser detuning was varied.  However, it was observed that, for a given laser detuning from the (zero-field) $^3P_1$ level, the spectra for the $m=+1$ and -1 states were essentially identical, and identical to that seen when exciting a degenerate mix of $m=0$ and $\pm1$ states in zero magnetic field.  These data clearly demonstrate that it is the intermediate state detuning that governs the spectral profiles.

\begin{figure}
    \centering
    \includegraphics[width=0.5\linewidth]{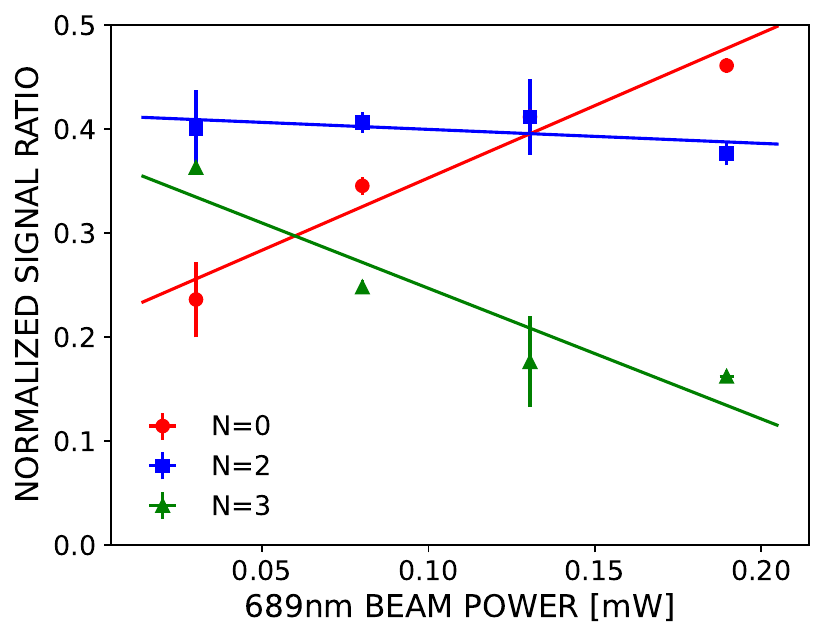}
    \caption{\label{fig:relative}
    The fractional contributions of the $N=0$, 2, and 3 features to the total $^{86}$Sr dimer spectra as a function of laser power for an intermediate state detuning of 15~MHz.
    }
\end{figure}

One possible explanation to consider is that the oscillations in the $N=0$ population result from variations in the scattering length, $a_s$.  To observe near-complete suppression of $N=0$ production, however, the scattering length would have to increase to $\sim1200$~a.u..
While an optical Feshbach resonance
~\cite{nbb15,ydr13} could lead to variations in the $^{86}$ULRRM spectrum, the observed behavior is not consistent with this mechanism.
Studies of the dependence of the $^{86}$Sr dimer spectral profiles on 689~nm laser power were also undertaken.  As shown in Fig.~\ref{fig:relative}
the relative size of the $N=0$ feature increases steadily with increasing laser power, whereas those of the $N=2$ and 3 features decrease, this decrease being particularly pronounced for the $N=3$ feature.  

Another possible explanation for the present observations is that the oscillatory $N=0$ feature results from the presence of some accidental resonance and that the second two features result from production of $N=0$ and 2 states.  While their separations are consistent with such a hypothesis and their relative sizes and behavior more consistent with model predictions, the measured isotope shifts speak against this explanation.  It should also be noted that the nearest Rydberg state to the $5s29s~^3S_1$ state, the $5s30d^1D_2$ state, lies $\sim20$~GHz away.  Furthermore, a two-electron calculation shows that there are no doubly-excited states nearby and that the first vibrationally-excited state is 30~MHz away.  The low sample density limits any production of trimers.

Another puzzle present in the $^{86}$Sr spectra is the creation of significant numbers of $N=3$ states which theory predicts should only be created in significant numbers for temperatures $\gtrsim10~\mu$K.  (Indeed, simple classical arguments suggest that for any reasonable choice of relative initial atom velocities and impact parameters, the initial mechanical angular momentum in the system is insufficient to allow creation of significant numbers of $N=3$ states.)  This suggests that, in the special case of $^{86}$Sr triplet dimers, some form of coupling between spin and rotational angular momentum must be present.  However, model calculations that include magnetic interactions indicate that these should have very little effect on the predicted results, and the origin of the $N=3$ production therefore remains a mystery.
\section{Conclusions}
The present measurements show that, whereas the behavior observed the production of $^{84}$Sr and $^{86}$Sr singlet dimers and $^{84}$Sr triplet dimers matches well that predicted using a theoretical model that incorporates the effects of the initial atom-atom scattering, sample temperature, and photon momentum transfer, the behavior seen in the production of $^{86}$Sr triplet dimers does not.  This indicates that, in this special case, effects are important that are not included in the present model and lead to changes in the relative production rates for the different rotational levels and the creation of $N=3$ rotational states.  However, no definitive mechanism is, as yet, identified.  The data therefore provide a puzzle that remains to be solved.

\begin{acknowledgments}
    Research supported by the NSF under Grants No. PHY 1904294 and PHY 2110596.  The calculations were undertaken using the Vienna cluster.
\end{acknowledgments}
\bibliography{bibliography}

\end{document}